# Secure Routing for Mobile Ad hoc Networks

**Panagiotis Papadimitratos and Zygmunt J. Haas**
**Wireless Networks Laboratory,**
**School of Electrical and Computer Engineering, Cornell University,**
**395 and 323 F.T. Rhodes Hall, Ithaca NY 14853**
**{papadp, haas}@ece.cornell.edu**

## Abstract
The emergence of the Mobile Ad Hoc Networking (*MANET*) technology advocates self-organized wireless interconnection of communication devices that would either extend or operate in concert with the wired networking infrastructure or, possibly, evolve to autonomous networks. In either case, the proliferation of *MANET*-based applications depends on a multitude of factors, with trustworthiness being one of the primary challenges to be met. Despite the existence of well-known security mechanisms, additional vulnerabilities and features pertinent to this new networking paradigm might render such traditional solutions inapplicable. In particular, the absence of a central authorization facility in an open and distributed communication environment is a major challenge, especially due to the need for cooperative network operation. In particular, in *MANET*, any node may compromise the routing protocol functionality by disrupting the route discovery process. In this paper, we present a route discovery protocol that mitigates the detrimental effects of such malicious behavior, as to provide correct connectivity information. Our protocol guarantees that fabricated, compromised, or replayed route replies would either be rejected or never reach back the querying node. Furthermore, the protocol responsiveness is safeguarded under different types of attacks that exploit the routing protocol itself. The sole requirement of the proposed scheme is the existence of a security association between the node initiating the query and the sought destination. Specifically, no assumption is made regarding the intermediate nodes, which may exhibit arbitrary and malicious behavior. The scheme is robust in the presence of a number of non-colluding nodes, and provides accurate routing information in a timely manner.

## A. INTRODUCTION
The provision of security services in the *MANET* context faces a set of challenges specific to this new technology. The insecurity of the wireless links, energy constraints, relatively poor physical protection of nodes in a hostile environment, and the vulnerability of statically configured security schemes have been identified [4,5] in literature as such challenges. Nevertheless, the single most important feature that differentiates *MANET* is the absence of a fixed infrastructure. No part of the network is dedicated to support individually any specific network functionality, with routing (topology discovery, data forwarding) being the most prominent example. Additional examples of functions that cannot rely on a central service, and which are also of high relevance to this work, are naming services, certification authorities (*CA*), directory and other administrative services.

Even if such services were assumed, their availability would not be guaranteed, either due to the dynamically changing topology that could easily result in a partitioned network, or due to congested links close to the node acting as a server. Furthermore, performance issues such as delay constraints on acquiring responses from the assumed infrastructure would pose an additional challenge.

The absence of infrastructure and the consequent absence of authorization facilities impede the usual practice of establishing a line of defense, separating nodes into trusted and non-trusted. Such a distinction would have been based on a security policy, the possession of the necessary credentials and the ability for nodes to validate them. In the *MANET* context, there may be no ground for an *a priori* classification, since all nodes are required to cooperate in supporting the network operation, while no prior security association can be assumed for all the network nodes. Additionally, in *MANET* freely roaming nodes form transient associations with their neighbors, join and leave *MANET* sub-domains independently and without notice. Thus it may be difficult in most cases to have a clear picture of the ad hoc network membership. Consequently, especially in the case of a large-size network, no form of established trust relationships among the majority of nodes could be assumed.

In such an environment, there is no guarantee that a path between two nodes would be free of malicious nodes, which would not comply with the employed protocol and attempt to harm the network operation. The mechanisms currently incorporated in *MANET* routing protocols cannot cope with disruptions due to malicious behavior. For example, any node could claim that is one hop away from the sought destination, causing all routes to the destination to pass through itself. Alternatively, a malicious node could corrupt

This work has been sponsored in part by the NSF grant number ANI-9980521 and the ONR contract number N00014-00-1-0564.

any in-transit route request (reply) packet and cause data to be misrouted.

The presence of even a small number of adversarial nodes could result in repeatedly compromised routes, and, as a result, the network nodes would have to rely on cycles of time-out and new route discoveries to communicate. This would incur arbitrary delays before the establishment of a non-corrupted path, while successive broadcasts of route requests would impose excessive transmission overhead. In particular, intentionally falsified routing messages would result in a *denial-of-service (DoS)* experienced by the end nodes. The proposed here scheme combats such types of misbehavior and safeguards the acquisition of topological information.

Our scheme guarantees that a node initiating a route discovery will be able to identify and discard replies providing false topological information, or, avoid receiving them. Our protocol departs from the Internet related solutions [2], which require the existence of a trust structure that encompasses all nodes participating in routing, and may rely on network management operations to detect routing instabilities. Moreover, the novelty of our scheme, as compared with other *MANET* secure routing schemes, is that false route replies, as a result of malicious node behavior, are discarded partially by benign nodes while in-transit towards the querying node, or deemed invalid upon reception. Most importantly, the above-mentioned goals are achieved with the existence of a security association between the pair of end nodes *only*, without the need for intermediate nodes to cryptographically validate control traffic.

The widely accepted technique in the *MANET* context of route discovery based on broadcasting query packets is the basis of our protocol. More specifically, as query packets traverse the network, the relaying intermediate nodes append their identifier (e.g., *IP* address) in the query packet header. When one or more queries arrive at the sought destination, replies that contain the accumulated routes are returned to the querying node; the source then may use one or more of these routes to forward its data. Reliance on this basic route query broadcasting mechanism allows our proposed here *Secure Routing Protocol (SRP)* to be applied as an extension of a multitude of existing routing protocols. In particular, the *Dynamic Source Routing* (*DSR*) [8] and the *IERP* [13] of the *Zone Routing Protocol* (*ZRP*) [14] framework are two protocols that can be extended in a natural way to incorporate *SRP*. Furthermore, other protocols such as *ABR* [15] for example, could be combined with *SRP* with minimal modifications to achieve the security goals of the *SRP* protocol.

*SRP* guarantees the acquisition of *correct* topological information in a timely manner, i.e., the route replies that are validated and accepted by the querying node provide accurate connectivity information, despite the presence of strong adversaries. The protocol is proven robust against a set of attacks that attempt to compromise the route discovery, under the assumption of *non-colluding* adversarial nodes.

In the sequel, we review schemes related to the problem at hand and then present our scheme. First, a concise overview is provided, followed by the detailed definition in section D. The analysis of the protocol is given next, with a discussion of related issues in section F.

## B. RELATED WORK

Outside the *MANET* community, secure routing in the Internet has, of course, received increased attention [2]. The proposed solutions rely mainly on the existence of a line of defense, separating the fixed routing infrastructure from all other network entities. This is achieved by distributing a set of public keys/certificates, which signify the authority of the router to act within the limits of the employed protocol (e.g., advertise certain routes), and allow all routing data exchanges to be authenticated, non-repudiated and protected from tampering. However, such approaches cannot combat a single malicious router disseminating incorrect topological information. More importantly, they are not applicable in the *MANET* context, because of impediments such as the absence of a fixed infrastructure and a central entity.

Despite the fact that security of *MANET* routing protocols is envisioned to be a major “roadblock” in commercial application of this technology, only a limited number of works has been published in this area. Such efforts have mostly concentrated on the aspect of data forwarding, disregarding the aspect of topology discovery. On the other hand, solutions that target route discovery have been based on approaches for fixed-infrastructure networks, defying the particular *MANET* challenges.

For the problem of secure data forwarding, two mechanisms that (i) detect *misbehaving* nodes and report such events and (ii) maintain a set of metrics reflecting the past behavior of other nodes [23] have been proposed to alleviate the detrimental effects of packet dropping. Each node may choose the ‘best’ route, comprised of relatively well-behaved nodes; i.e., nodes that do not have history of avoiding forwarding packets along established routes. Among the assumptions for the above-mentioned work are a shared medium, bi-directional links, use of source routing (i.e., packets carry the entire route that becomes known to all intermediate nodes), and no colluding malicious nodes. Nodes operating in promiscuous mode overhear the transmissions of their successors and may verify whether the packet was forwarded to the downstream node and check the integrity of the forwarded packet. Upon detection of a misbehaving node, a report is generated and nodes update the rating of the reported misbehaving node. The ratings of nodes along a well-behaved route are periodically incremented, while reception of a misbehavior alert dramatically decreases the node rating.[1] When a new route is required, the source node calculates a path metric equal to the average of the ratings of the nodes in each of the route replies, and selects the route with the highest metric.

The detection mechanism exploits two features that frequently appear in *MANET:* the use of a shared channel and source routing. Nevertheless, the plausibility of this solution could be questioned for several reasons and, indeed, the authors provide a short list of scenarios of incorrect detection. The possibility of falsely detecting misbehaving nodes could easily create a situation with many nodes falsely suspected for a long period of time. In addition, the metric construction may lead to a route choice that includes a suspected node, if, for example, the number of hops is relatively high, so that a low rating is "averaged out." Finally, the most important vulnerability is the proposed feedback itself; there is no way for the source, or any other node that receives a misbehavior report to validate its authenticity or correctness. Consequently, the simplest attack would be to generate fake alerts and eventually disable the network operation altogether. The protocol attempts new route discoveries when none of the route replies is free of suspected nodes, with the excessive route request traffic degrading the network performance. At the same time, the adversary can falsely accuse a significant fraction of nodes within the time-out period related to reinstating from a negative rating and, essentially, partition the network.

A different approach [24] is to provide incentive to nodes, so that they comply with protocol rules; i.e., properly relay user data. The concept of fictitious currency is introduced, in order to *endogenize* the behavior of the assumed greedy nodes, which would forward packets in exchange for currency. Each intermediate node purchases from its predecessor the received data packet and sells it to its successor along the path to the destination. Eventually the destination pays for the received packet.[2] This scheme assumes the existence of an overlaid geographic routing infrastructure and a *Public Key Infrastructure (PKI).* All nodes are pre-loaded with an amount of currency, have unique identifiers, are associated with a pair of private/public keys and all cryptographic operations related to the currency transfers are performed by a physically tamper-resistant module. The applicability of the scheme, which targets wide-area *MANET*, is limited by the assumption of an on-line Certification Authority in the *MANET* context. Moreover, nodes could flood the network with packets destined to non-existent nodes and possibly lead nodes unable to forward purchased packets to starvation. The practicality of the scheme is also limited by its assumptions, the high computational overhead (hop-by-hop public key cryptography, for each transmitted packet), and the implementation of physically tamper-resistant modules.

The protection of the route discovery process has been regarded as an additional Quality-of-Service (*QoS*) issue [17], by choosing routes that satisfy certain quantifiable security criteria. In particular, nodes in a *MANET* subnet are classified into different trust and privilege levels. A node initiating a route discovery sets the sought security level for the route; i.e., the required minimal trust level for nodes participating in the query/reply propagation. Nodes at each trust level share symmetric encryption and decryption keys. Intermediate nodes of different levels cannot decrypt in-transit routing packets, or determine whether the required *QoS* parameter can be satisfied, and simply drop them. Although this scheme provides protection (e.g., integrity) of the routing protocol traffic, it does not eliminate false routing information provided by malicious nodes. Moreover, the proposed use of symmetric cryptography allows any node to corrupt the routing protocol operation within a level of trust, by mounting virtually any attack that would be possible without the presence of the scheme. Finally, the assumed supervising organization and the fixed assignment of trust levels does not pertain to the *MANET* paradigm. In essence, the proposed solution transcribes the problem of secure routing in a context where nodes of a certain group are assumed to be trustworthy, without actually addressing the global secure routing problem.

An extension of the *Ad Hoc On-demand Distance Vector* (*AODV*) [16] routing protocol has been proposed [18] to protect the routing protocol messages. The *Secure-AODV* scheme assumes that each node has certified public keys of all network nodes, so that intermediate nodes can validate all in-transit routing packets. The basic idea is that the originator of a control message appends an *RSA signature* [19] and the last element of a *hash chain* [20] (i.e., the result of *n* consecutive hash calculations on a random number). As the message traverses the network, intermediate nodes cryptographically validate the signature and the hash value, generate the *k-th* element of the hash chain, with *k* being the number of traversed hops, and place it in the packet. The route replies are provided either by the destination or intermediate nodes having an active route to the sought destination, with the latter mode of operation enabled by a different type of control packets.

The use of public-key cryptography imposes a high processing overhead on the intermediate nodes and can be considered unrealistic for a wide range of network instances. Furthermore, it is possible for intermediate nodes to corrupt the route discovery by pretending that the destination is their immediate neighbor, advertising arbitrarily high sequence numbers and altering (either decreasing by one or arbitrarily increasing) the actual route length. Additional vulnerabilities stem from the fact that the *IP* portion of the *S-AODV* traffic can be trivially compromised, since it is not (and cannot be, due to the *AODV* operation) protected, unless additional hop-by-hop cryptography and accumulation of signatures is used. Finally, the assumption that certificates are bound with *IP* addresses is unrealistic; roaming nodes joining *MANET* sub-domains will be assigned *IP* addresses dynamically (e.g., *DHCP* [21]) or even randomly (e.g., *Zero-Configuration* [22]).

[1] The initial rating, 0.5, is increased by 0.01 every 200 ms. Suspected nodes have a rating equal to –100, with the option for a long timeout period after which the negative rating is changed back to a positive value.

[2] An alternative implementation, with each packet carrying a purse of fictitious currency from which nodes remove their reward, faces different challenges as well.

A different approach is taken by the *Secure Message Transmission (SMT)* [1] protocol, which, given a topology view of the network, determines a set of diverse paths connecting the source and the destination nodes. Then, it introduces limited transmission redundancy across the paths, by dispersing a message into *N* pieces, so that successful reception of any *M*-out-of-*N* pieces allows the reconstruction of the original message at the destination. Each piece, equipped with a cryptographic header that provides integrity and replay protection along with origin authentication and is transmitted over one of the paths. Upon reception of a number of pieces, the destination generates an acknowledgement informing the source of which pieces, and thus routes, were intact. In order to enhance the robustness of the feedback mechanism, the small-sized acknowledgments are maximally dispersed (i.e., successful reception of at least one piece is sufficient) and are protected by the protocol header as well. If less than *M* pieces were received, the source re-transmits the remaining pieces over the intact routes. If too few pieces were acknowledged or too many messages remain outstanding, the protocol adapts its operation, by determining a different path set, re-encoding undelivered messages and re-allocating pieces over the path set. Otherwise, it proceeds with subsequent message transmissions.

The protocol exploits *MANET* features such as the topological redundancy, interoperates widely with accepted techniques such as source routing, relies on a security association between the source and the destination, and makes use of highly efficient symmetric-key cryptography. It does not impose processing overhead on intermediate nodes, while the end nodes make the routing decisions, based on the feedback provided by the destination and the underlying topology discovery and route maintenance protocols. The fault-tolerance of *SMT* is enhanced by the adaptation of parameters such as the number of paths and the dispersion factor (i.e., the ratio of required pieces to the total number of pieces). *SMT* can yield *100%* successful message reception, even if *10* to *20* percent of the network nodes are malicious. Moreover, algorithms for the selection of path sets with different properties, based on different metrics and the network feedback, can be implemented by *SMT*. *SMT* provides a flexible, end-to-end, secure traffic engineering scheme tailored to the *MANET* characteristics.

It is noteworthy that *SMT* provides a limited protection against the use of compromised topological information, although its main focus is to safeguard the data forwarding operation. The use of multiple routes compensates for the use of partially incorrect routing information [4], rendering a compromised route equivalent to a route failure. Nevertheless, the disruption of the route discovery can still be the most effective way for adversaries to consistently compromise the communication of one or more pairs of nodes. This is where *SRP* can complement *SMT.*

*SRP* safeguards the route discovery and makes use of cryptographic tools, an indispensable requirement for any security scheme. Only the end nodes have to be securely associated, and there is no need for cryptographic validation of control traffic at intermediate nodes, two factors that render the scheme efficient and scalable. *SRP* places the overhead on the end nodes, an appropriate choice for a highly decentralized environment, and contributes to the robustness and flexibility of the scheme. Moreover, *SRP* does not rely on state stored in intermediate nodes, thus is immune to malicious acts not directed against the nodes that wish to communicate in a secure manner. Finally, *SRP* provides one or more route replies, whose correctness is verified by the route "geometry" itself. A querying node acquires correct network connectivity information and the ability to choose an optimal route, with respect to the number of hops or another criterion. At the same time, the overall routing and control traffic overhead under highly adverse conditions is reduced, and protection of the end nodes against attacks that aim at exhausting their resources, is provided.

## C. THE PROPOSED SCHEME

### C.1. Basic Assumptions

We focus on bi-directional communication between a pair of nodes. A *security association (SA)* between the *source node S* and the *destination node T* is assumed. The trust relationship could be instantiated, for example, by the knowledge of the public key of the other communicating end. The two nodes can negotiate a shared secret key, e.g., via the Elliptic Curve Diffie-Hellman algorithm [7,12], and then, using the *SA*, verify that the principal that participated in the exchange was indeed the trusted node. For the rest of the discussion, we assume the existence of a shared key $K_{S,T}$. The *SA* is bi-directional in that the shared key can be used for control (data) traffic flow in both directions. Relevant state has to be maintained for each direction though.

The existence of the *SA* is justified, because the end hosts chose to employ a secure communication scheme and,

consequently, should be able to authenticate each other. For example, such a group (pair) of nodes could have performed a secure key exchange [10], or an initial distribution of credentials. However, the existence of *SA*'s with any of the intermediate nodes is unnecessary. Finally, it is required that end nodes be able to use static or non-volatile memory.

The adversarial nodes may attempt to compromise the network operation by exhibiting arbitrary, *Byzantine* behavior [3]. They are able to corrupt, replay, and fabricate routing packets. They may attempt to misroute them in any possible manner and, in general, they cannot be expected to properly execute the routing protocol. Although a set of malicious nodes may mount attacks against the protocol concurrently, we assume that nodes are not capable of colluding within one step of the protocol execution; that is, within the period of broadcasting one query and reception of the corresponding replies. For clarification, we discuss below an attack mounted by two colluding nodes during a single route discovery.

The underlying data link layer (e.g., *IEEE 802.11* [6]) provides reliable transmission on a link basis, without any requirement of data link security services, such as the Wired Equivalent Protocol (*WEP*) function. Moreover, links are assumed to be bi-directional, a requirement fulfilled by most of the proposed medium access control protocols, especially the ones employing the *RTS/CTS* dialogue. It is also expected that a one-to-one mapping between *Medium Access Control* and *IP* addresses exists. Finally, the broadcast nature of the radio channel mandates that each transmission is received by all neighbors, which are assumed to operate in promiscuous mode.[3]

## C.2. Overview

Our work provides a novel approach to the secure route discovery operation for *MANET* routing protocols. The proposed here scheme combats attacks that disrupt the route discovery process and guarantees, under the above-mentioned assumptions, the acquisition of correct topological information. It also incorporates mechanisms that safeguard the network functionality from attacks exploiting the protocol itself, in order to degrade network performance and possibly lead to denial of service.

The source node *S* initiates the route discovery, by constructing a route request packet identified by a pair of identifiers: a query sequence number and a random query identifier. The source and destination and the unique (with respect to the pair of end nodes) query identifiers are the input for the calculation of the *Message Authentication Code (MAC)* [9], along with $K_{S,T}$. In addition, the identities (*IP* addresses) of the traversed intermediate nodes are accumulated in the route request packet.

Intermediate nodes relay route requests, so that one or more query packets arrive at the destination, and maintain a limited amount of state information regarding the relayed queries, so that previously seen route requests are discarded. Moreover, they provide feedback in the event of a path breakage, and in some cases they may provide route replies, as explained in section D.5.

The route requests reach the destination *T*, which constructs the route replies; it calculates a *MAC* covering the route reply contents and returns the packet to *S* over the *reverse* of the route accumulated in the respective request packet. The destination responds to one or more request packets of the same query, so that it provides the source with an as diverse topology picture as possible.[4] The querying node validates the replies and updates its topology view.

As an illustrative example, consider the topology of Fig.1, comprising ten nodes. *S* queries the network to discover one or more routes to *T*. The nodes $M_1$ and $M_2$ are two malicious intermediate nodes. We denote the query request as a list $\{Q_{S,T};n_1,n_2,..,n_k\}$, with $Q_{S,T}$ denoting the *SRP* header for a query searching for *T* and initiated by *S*. The $n_i$, $i\neq\{1,k\}$, are the *IP* addresses of the traversed intermediate nodes and $n_1=S$, $n_k=T$. Similarly, the route reply is denoted as $\{R_{S,T};n_1,n_2,..,n_k\}$. We now consider a number of scenarios of possible security attacks by the two malicious nodes.

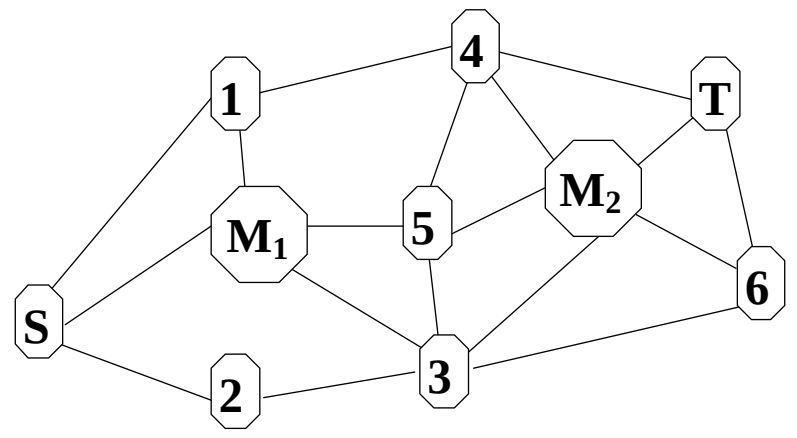

**Figure 1.** Example Topology: *S* wishes to discover a route to *T* in the presence of two malicious nodes, $M_1$ and $M_2$.

Scenario 1: Consider the case that when $M_1$ receives $\{Q_{S,T};S\}$, it attempts to mislead *S* by generating $\{R_{S,T};S,M_1,T\}$. Not only would *S* accept such a reply, if a regular routing protocol were used, but it would most probably choose this fake route, since $\{S,M_1,T\}$ would have fewer hops than any other legitimate reply. It would also be received with the least delay, because of the close distance between $M_1$ and *S*. The requirement that the request reaches the destination disallows any intermediate node to provide a reply in this manner, and, the false reply packet is discarded, since $M_1$ does not possess $K_{S,T}$ and cannot generate a valid *MAC*.

[3] I.e., able to overhear all transmissions from nodes within the range of their transceiver

[4] The number of replies and the time-window the destination allocates for replies to a specific query are design parameters. Moreover, the source could provide an indicator of the required diversity, so that *T* can regulate the number of replies.

Scenario 2: Consider the case in which $M_1$ discards request packets arriving from its neighbors, excluding the one from node *1*. This type of malicious act cannot be countered, but the controlled flooding of the query packets provides the required robustness. By discarding route request packets, a malicious node partially narrows the topology view of *S* and, to some extend, impedes the network operation. In essence, the malicious node can always hide its incident links, but at the same time it practically removes itself from *S*'s view. Thus, it cannot inflict harm to data flows originating from *S*, since the routes chosen by *S* would simply exclude $M_1$.

Scenario 3: As assumed above, $M_1$ sees and appropriately relays $\{Q_{S,T};S,1,M_1\}$; upon arrival of $\{Q_{S,T};S,1,M_1,5,4\}$ at *T*, the reply is generated and routed over the reverse path. When $M_1$ receives $\{R_{S,T};S,1,M_1,5,4,T\}$, it tampers with its content and relays $\{R_{S,T};S,1,M_1,Y,T\}$, with *Y* being any invented sequence of nodes. *S* readily discards the reply, due to the integrity protection provided by the *MAC*.

Scenario 4: When $M_2$ receives $\{Q_{S,T};S,2,3\}$, it corrupts the accumulated route and relays $\{Q_{S,T};S,X,3,M_2\}$ to its neighbors, where *X* is a false, invented *IP* address (or, any sequence of *IP* addresses). This request arrives at *T*, which constructs the reply and routes it over $\{T,M_2,3,X,S\}$ towards *S*. When node *3* receives the reply, it cannot forward it any further, since *X* is not its neighbor, and the reply is dropped.

Scenario 5: In order to consume network resources, $M_1$ replays route requests, which are discarded by intermediate nodes, since they maintain a list of query identifiers seen in the past. This is achieved by the underlying routing protocol itself, within the limitations imposed by the size of the query table. But queries replayed after a significant period of time, will propagate across the network and arrive at *T*. The *query sequence number*, used only by the end nodes for the query identification, allows *T* to discard such queries. If the request header were corrupted, the query would also be discarded. Similarly, *T* discards fabricated route requests, since malicious nodes cannot generate valid request *MAC*.

Scenario 6: Assume that $M_1$, after observing a few route requests originating from *S*, fabricates several queries with the subsequent query identifiers. The goal of this attack is to make intermediate nodes store these identifiers and discard legitimate, future $\{Q_{S,T};n_1,\ldots,n_j\}$ route requests. The cost of this attack is low (a single route request transmission per identifier) and, with the *Time-To-Live (TTL)* field of the query packet set to a high value, the affected network area may be significantly large. The query identifier values used by intermediate nodes implementing *SRP* are 'unique' and random, unlike the query identification field of existing on-demand routing protocols, whose values are a monotonically increasing sequence. Consequently, such an attack cannot practically affect the protocol operation, because of the extremely low probability of predicting the query identifiers.

Scenario 7: Node $M_1$ attempts to forward $\{Q_{S,T};S,M^*\}$; i.e., it *spoofs* an *IP* address. Such an act is possible and at the routing protocol level the query would propagate through the network and reach *T*. Consequently, S would accept $\{R_{S,T};S,M^*,1,4,T\}$ as a route. It is apparent that the connectivity information conveyed by such a reply is correct. Indeed, all that $M_1$ would achieve is to mask its identity, which in general will be temporary. Thus, the malicious node would not achieve anything more than its placement on a potential $S \rightarrow T$ route, which would have been possible in the first place, without any *IP* spoofing.

Scenario 8: Now, let us assume that $M_1$ attempts to return a number of replies, each with a different spoofed *IP* address, namely, $M_i,M_{i+1},\ldots M_{i+j}$, i.e., an "extension" of Scenario 7. This would lead *S* to believe that a multitude of possible routes to *T* exist, although, in reality, all of these routes are controlled by $M_1$. As explained in Scenario 1, $M_1$ is not allowed to generate replies, and thus fabricate ones that contain the spoofed addresses. An alternative way for $M_1$ to mount this attack would be to relay more than one route requests, placing a different *IP* address in each of them; *T* would generate the corresponding replies, $M_1$ would relay them back towards the source, and *S* would have no choice but to accept them. Fortunately, such an attack is successfully countered by our protocol: $M_1$'s neighbors relay only one route request, with specific source and target nodes and query identifier. For example, nodes *1,3* and *5* will relay the first of such queries and drop subsequent packets as previously seen requests, thanks to the broadcast channel. If $M_1$ modified the query identifier, the forged query would be forwarded, but *T* would detect the alteration, due to the *MAC*, and drop the request.

The only possible attack against the protocol would be if nodes colluded during the two phases of a single route discovery. In such a case, they would manage to make the source node to accept partially false routing information. For example, in Fig.1, when $M_1$ receives the route request, it can tunnel it to $M_2$; i.e. discover a route to $M_2$ and send the request encapsulated in a data packet. Then, $M_2$ broadcasts a request with the route segment between $M_1$ and $M_2$ falsified, e.g. $\{Q_{S,T};S,M_1,Z,M_2\}$. *T* receives the request and constructs a reply, which is routed over $\{T,M_2,Z,M_1,S\}$. $M_2$ receives the reply and tunnels it back to $M_1$, which, then, returns it to *S*. As a result, the connectivity information is only partially correct (in this example, only the first and last link). However, one pair of colluding nodes can convince *S* of only a single false path that will include the two nodes. The reason is that $M_2$ cannot forward a number of requests towards *T* using spoofed *IP* addresses, as explained above. Special care is needed for a case similar to Fig.1, where $M_2$ is adjacent to *T*, with countermeasures discussed in the sequel. In particular, the application of *SMT* on top of the proposed here secure routing protocol can further mitigate the impact of an attack mounted by colluding nodes.

## D. DETAILED PROTOCOL DESCRIPTION

The *Secure Routing Protocol (SRP)* introduces a set of new features that can be incorporated in the context of the underlying *basis protocol* with low overhead. In principle, it can retain mechanisms, such as the control of the query propagation, the rate of query generation, and the neighbor discovery protocol, if present. *SRP* extends the basic protocol by enforcing rules on the format and propagation of route request, route reply, and the error messages, by introducing the required additional functionality.

In short, *SRP* makes efficient use of the security association between the two communicating nodes *S* and *T*. Route request packets verifiably propagate to the destination (in the general case) and route replies are returned to *S* strictly over the reversed route, as accumulated in the route request packet. Similarly, route error messages can only be generated by nodes that lie on the route that is reported as broken. In order to guarantee this functionality of crucial importance, *SRP* determines explicitly the interaction with the network layer; i.e., the *IP*-related functionality. Furthermore, it provides a novel way of query identification, which protects the query propagation and the end-nodes from *DoS* attacks. Finally, propagating query packets are handled locally by a *priority scheme* that enhances the robustness and the responsiveness of the protocol.

0 1 2 3
0 1 2 3 4 5 6 7 8 9 0 1 2 3 4 5 6 7 8 9 0 1 2 3 4 5 6 7 8 9 0 1

| |
|---|
| IP Header |
| Basis Routing Protocol Packet |
| SRP Header |

**Figure 2.** *SRP* as an extension of a reactive routing protocol: the *SRP* header (shown in detail in Fig. 3) is appended to the basis routing protocol header (shaded area).

The features introduced by *SRP* require the addition of a 6-word header, as shown in Fig.3. The *SRP Header* is integrated into the underlying protocol header structure as an additional *IP option* (Fig.2), and covers most parts of the routing protocol datagram. Different types of SRP messages are distinguished with the help of the 1-byte *Type* field. In this work, we primarily consider the augmentation of route request and reply packets and in the sequel each message type is described individually. However, it is possible for *SRP* to operate in a more general setting, where, for example, a route reply is appended to a data packet.

### D.1. Route Request

A source node *S* maintains a *Query Sequence number* $Q_{seq}$ for each destination it securely communicates with. This 32-bit sequence number increases monotonically, for each route request generated by *S*, and allows *T* to detect outdated route requests. The sequence number is initialized at the establishment of the *SA* and although it is not allowed to wrap around, it provides approximately a space of four billion query requests per destination. If the entire space is used, a new security association has to be established, in one of the ways described in C.1.

0 1 2 3
0 1 2 3 4 5 6 7 8 9 0 1 2 3 4 5 6 7 8 9 0 1 2 3 4 5 6 7 8 9 0 1

| | |
|---|---|
| Type | Reserved |
| Query Identifier | |
| Query Sequence Number | |
| SRP MAC | |

**Figure 3.** *SRP* Header: for a pair of source and destination nodes, the control message, identified by its *Type*, is uniquely distinguished by the pair of identifiers. The Message Authentication Code *(MAC)* covers parts of the message, depending on its type.

For each outgoing *Route Request*, *S* generates a 32-bit random *Query Identifier* $Q_{ID}$, which is used by intermediate nodes as a means to identify the request. $Q_{ID}$ is the output of a secure pseudorandom number generator [11]; its output is statistically indistinguishable from a truly random one and is unpredictable by an adversary with limited computational power. Since intermediate nodes have limited memory of past queries, uniqueness and randomness can be efficiently achieved, by using a one-way function (e.g., *SHA-1* [27]) and a small random seed as input. This renders the prediction of the query identifiers practically impossible, and combats an attack where malicious nodes simply broadcast fabricated requests only to cause subsequent legitimate queries to be dropped.

Both $Q_{ID}$ and $Q_{seq}$ are placed in the *SRP* header, along with appropriate *Type* value and the Request *Message Authentication Code (MAC)*. The *MAC* is a 96-bit long field, generated by a keyed hash algorithm [9], which calculates the truncated output of a one-way or hash function (e.g., *SHA-1* or *MD5* [28]). The one-way function input is the entire *IP* header, the basis protocol route request packet and most importantly, the shared key $K_{S,T}$. The *Route Request* fields that are updated as the packet propagates towards the destination, i.e., the accumulated addresses of the intermediate nodes, and the *IP*-header mutable fields are excluded.

## D.2. Query Handling/Propagation

Intermediate nodes parse the received Route Requests in order to determine whether an *SRP* header is present. If not, they process the packet as described in the basis protocol specification. Otherwise, the intermediate nodes extract the $Q_{ID}$. The source and the destination addresses are also extracted in order to create an entry in the *query table*. Queries with $Q_{ID}$ matching one of the table entries for the same pair of end nodes are discarded. Otherwise, the intermediate nodes re-broadcast the route request.

Intermediate nodes also measure the frequency of queries received from their neighbors, in order to regulate the query propagation process. On one hand, all nodes self-regulate the generation of new route requests, in order to maintain the control traffic overhead low. On the other hand, malicious nodes probably act selfishly and avoid backing off before generating a new route query, or generate queries at the highest possible rate, consuming network resources and degrading the routing protocol performance.

In order to guarantee the responsiveness of the routing protocol, each benign node maintains a priority ranking of its neighbors according to the corresponding observed rate of queries. The highest priority is assigned to the nodes generating (or relaying) requests with the lowest rate, and the lowest priority to the neighbors that generate queries more frequently. Then, quanta are allocated proportionally to the priorities and within each class queries are serviced in a round-robin manner.

As immediate neighbors of a malicious node observe a high rate of incoming queries, they update the corresponding weight (priority). Moreover, not serviced low priority queries are eventually discarded. In this way, non-malicious queries are only affected for a time period equal to the time it takes to detect and update the priority assigned to a misbehaving neighbor. At the same time, the round-robin operation provides additional assurance that benign requests will propagate as well. More importantly, the filtering of the suspected requests will be performed close to the potential source of misbehavior, and benign nodes farther away from the adversary will not be affected, as they will have to relay fabricated queries at a lower rate.

## D.3. Route Reply

*T* validates the received route request packet, by first verifying that it has originated from a node with which it has a security binding. Then, $Q_{seq}$ is compared to $S_{max}$, the maximum query sequence number received from *S*, within the lifetime of the *SA*. If $Q_{seq} \leq S_{max}$, the request is discarded as outdated or replayed. Otherwise, *T* calculates the keyed hash of the request fields. If the output matches the *SRP* header *MAC*, the integrity of this request is verified, along with the authenticity of its origin.

The destination generates a number of replies to valid requests, at most as many as the number of its neighbors, in order to disallow a possibly malicious neighbor to control multiple replies. For each valid request, *T* places the accumulated route in the route reply packet and the $Q_{ID}$ and $Q_{seq}$ of the route request in the corresponding *SRP* header fields, so that *S* can verify the freshness of the reply. The *MAC* covers the basis protocol route reply and the rest of the *SRP* header, protects the integrity of the reply on its way to the source and offers evidence to *S* that the request has indeed reached the destination.

An alternative, more efficient implementation would be for the destination (*T*) to source-route a reply with an empty payload. The *SRP* header *Type* indicates that the packet is a reply, the source-route of the datagram contains the sought route reversed, and the *MAC* covers the *IP* source-route, as created by *T*. If the reply is deemed valid, *S* extracts the node sequence from the reply *IP* source-route and reverses it, in order to create the $S \rightarrow T$ route, or, simply decomposes it into its constituent links.

## D.4. Route Reply Validation

On reception of a *Route Reply, S* checks the source and destination addresses, $Q_{ID}$ and $Q_{seq}$ and discards the *Route Reply* if it does not correspond to the currently pending query. Otherwise, it compares the reply *IP* source-route with the reverse of the route carried in the reply payload. If the two routes match, *S* calculates the *MAC* using the replied route, the *SRP* header fields and $K_{S,T}$. Upon successful verification, *S* is assured that the request, indeed, reached *T* and that the reply was not corrupted on its way from *T* to *S*. Moreover, since the reply packet has been routed and successfully received over the reverse of the route it carries, the route information has not been compromised during the request propagation; i.e., before arriving at *T*. Thus, the connectivity information is genuine.

If the alternative form of reply with empty payload is returned, it is sufficient to validate the *MAC*, since the *IP* source-route provides the (reversed) route itself and implies that the reply arrived over this route. Moreover, if an intermediate node *V* having an *SA* with *S* provides a reply, the route suffix is accepted as genuine. In other words, *V* is trusted to provide a correct $V \rightarrow T$ route, and the above-mentioned checks are performed for the $S \rightarrow V$ route segment. If this is proven to be genuine, then the entire route is deemed genuine.

## D.5. Intermediate Node Replies

The caching of overheard routes is a severe vulnerability, since false topology information can be easily disseminated throughout a large portion of the network. A malicious node can fabricate data packets or route replies, which are, for example, cached by nodes operating in promiscuous mode. When such routes are used or provided as replies, more unsuspecting nodes cache such invalid routes and may use them in the future.

In order to achieve the required robustness, route caching is not encouraged in general and intermediate nodes are not required to provide route replies. However, we realize that route caching can improve the effectiveness of the route discovery process. As such, if an intermediate node *V* has an active route to *T* and an *SA* exists between *S* and *V*, then, a reply could be provided to *S*. This is the only case that the route request does not actually reach the destination.

This extension of the *SRP* functionality is enabled by the *Intermediate Node Reply Token (INRT)* (Fig. 4), and we propose here two alternative designs. Let $K_G$ be a *group key*, i.e., a secret shared by the members of a small group of nodes that *S* belongs to. At the same time, *S* and *T*, as every pair of the group nodes, have established an *SA*, i.e., share a secret key, as previously discussed. Then, *INRT* is merely the keyed hash of the route request message, calculated exactly as in D.1, apart from the fact that the key is $K_G$, instead of $K_{S,T}$. Any group node, namely *V*, that has an active route to *T* validates the request based on *INRT* and generates the reply, as described in D.3, using $K_{S,V}$. Alternatively, instead of extending the header, the source node could simply use $K_G$ for the *MAC* calculation. This would be a plausible solution only if *T* belonged to the group as well.

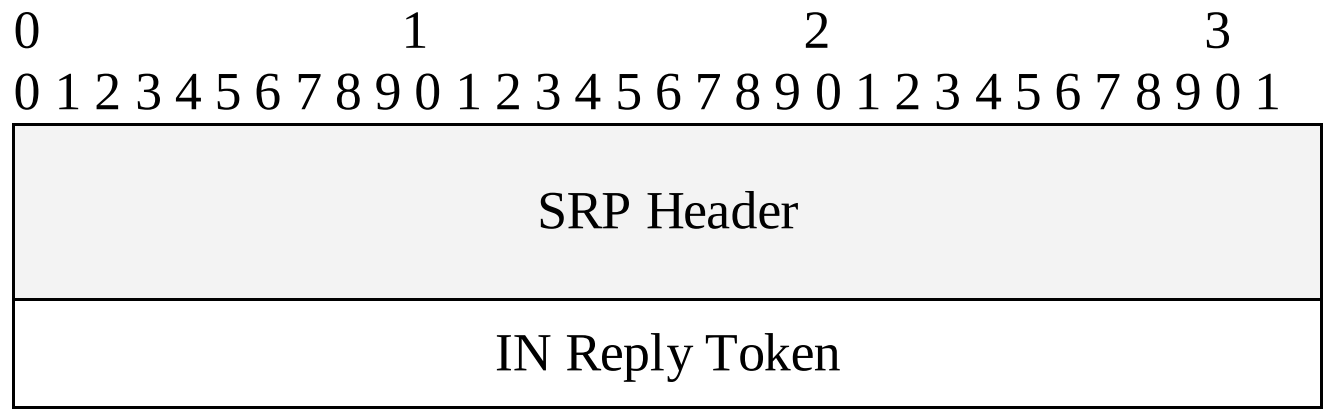

**Figure 4.** Extended *SRP* Header: the *INRT* allows intermediate group member nodes to validate the request and provide a reply.

A different method would be to calculate *INRT* as a digital signature; i.e., the hash of the route request encrypted with the private key of *S*. Then, any receiving node can validate the request and provide the reply. This mode would be useful in a scenario that a node does not belong to a group but still is securely associated with nodes other than *T*.

### D.6. Route Maintenance

This function, though not directly related to the route discovery, is an integral part of most *MANET* routing protocols. Topology changes have to be detected and the sources of the affected routes have to be notified, while avoiding false or fabricated notifications. This task is facilitated by the fact that intermediate node caching is disabled, but *route error* messages must be retained even if *SMT* is used in conjunction with *SRP*. The *SMT* acknowledgments allow for enhanced detection of any type of transmission failures. However, this end-to-end approach does not allow distinguishing benign (due to topology changes) from malicious route failures.

Thus, route error messages generated by intermediate nodes are retained in *SRP*, in order to provide fast detection of path breakages. The route error packets are source-routed along the prefix of the route reported as broken, and *S* compares the route traversed by the error message to the prefix of the corresponding route. In this way, it can verify that the provided route error feedback refers to the actual route and is not generated by a node that is not part of the route: The correctness of the feedback (i.e., whether it reports an actual failure to forward a packet) cannot be verified.

For example, in Fig.1, if the route *{S,1,4,T}* had been chosen, $M_2$ could simply generate a route error reporting the *(4,T)* link breakage, even though the route was intact. In order to get the error message to *S*, $M_2$ has to *source-route* it to *S*, and it does so over $\{M_2,4,1,S\}$ for example. Even though node *4* may not discard such a message[5], *S* will compare the source-route of the error message to the route reported as broken, or more specifically, the (reverse) segment reaching the broken link. The comparison fails, and the feedback is discarded, since *S* infers that an outlying node generated the route.

A malicious node lying on an $S \rightarrow T$ route can at most invalidate the route, mislead *S* by corrupting error messages generated by another node, or by masking a dropped packet as a link failure. Consequently, a malicious node can harm only the route it belongs, something that is possible in the first place if it simply dropped or corrupted data packets. On the other hand, it is important that under normal conditions the responsiveness of the protocol remains high.

## E. PROTOCOL CORRECTNESS PROOF

This section presents a formal analysis of the protocol and verifies that the stated goals are achieved. The analysis follows the methodology of [25]. Based on a set of assumptions, the current beliefs of the participating principals are derived from their initial beliefs and possessions. In particular, we follow the notation and inference rules in [26], and the Appendix provides a concise reference to the used notation.

[5] In fact, this is not required and it would pose additional processing overhead to intermediate nodes. On the other hand, node *4* could not accept to forward a packet fabricated by $M_2$, spoofing *4*'s address, in order to convince *S* that the route message originated from node *4*.

The protocol is abstracted as the exchange of two messages, a route request and a route reply. The messages are transmitted over a *public channel*; i.e., a sequence of intermediate nodes that may cause any impairment. The idealized form (i.e., the protocol with parts of the messages that do not contribute to the participants' beliefs omitted) is shown in Fig. 5.

$Q_{S,T}$ is the route request and $H$ is the Message Authentication Code *(MAC)* function. The relevant fields of $Q_{S,T}$ are the sequence number $Q_{seq}$, and the source and destination node addresses. As for the *route reply*, denoted as $R_{S,T}$, the $Q_{seq}$ field binds $R_{S,T}$ to the corresponding $Q_{S,T}$, and *route* is the actual route along which $T$ returns the reply.

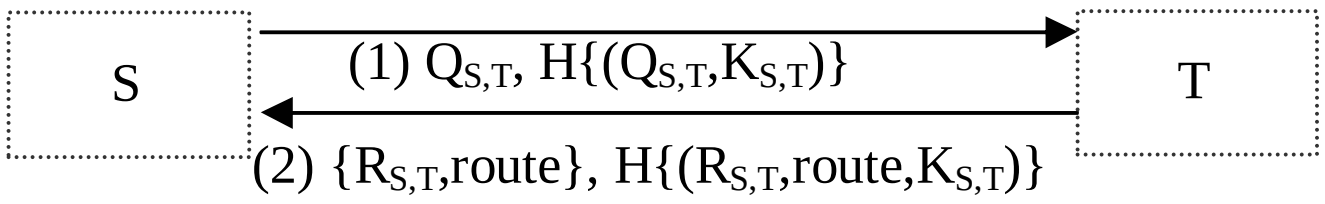

**Figure 5.** Idealized *SRP*: the protocol viewed as an exchange of two messages, without the fields that do not contribute to the participants' beliefs.

The initial assumptions are:

(i) $S \ni K_{S,T},\quad S \mid\equiv S \xleftrightarrow{K_{S,T}} T,\quad S \ni N_{S,T},\quad S \mid\equiv \#(N_{S,T})$

The sender possesses the shared key and it believes it is used for mutual proofs of identity between $S$ and $T$. It possesses $N_{S,T}$, the newly generated sequence number, and believes that $N_{S,T}$ has not been used before.

(ii) $T \ni K_{S,T},\quad T \mid\equiv S \xleftrightarrow{K_{S,T}} T,\quad T \ni N^{p}_{S,T}$

The receiver also trusts the shared secret, possesses the set of sequence numbers seen in the past and believes they were once uttered by $S$ ($T \triangleleft N^{p}_{S,T}, T \mid\equiv S \mid\sim N^{p}_{S,T}$). If the message (1) is the first transmission from $S$ to $T$ (within the lifetime of the *SA*) the set of past sequence numbers is merely the $Q_{seq}$ initialized at the *SA* establishment. Otherwise, the *SA* state justifies such a belief, which is our basis hypothesis. Moreover, $S$ and $T$ believe they are able to recognize $Q_{S,T}$ and $R_{S,T}$, respectively. [6]

(iii) $S \mid\equiv \otimes(R_{S,T}, H(R_{S,T}, K_{S,T})),\quad T \mid\equiv \otimes(Q_{S,T}, H(Q_{S,T}, K_{S,T}))$

For message (1), we have:

(iv) $\dfrac{T \triangleleft *(Q_{S,T}, H(Q_{S,T}, K_{S,T}))}{T \triangleleft (Q_{S,T}, H(Q_{S,T}, K_{S,T}))}$ and $\dfrac{T \triangleleft (Q_{S,T}, H(Q_{S,T}, K_{S,T}))}{T \ni (Q_{S,T}, H(Q_{S,T}, K_{S,T}))}$

i.e., $T$ sees a packet with the "not-originate-here" property, that is, it can distinguish, acting as a receiver, whether it has previously transmitted the packet in the current run.

(v) $\dfrac{T \ni (Q_{S,T}, H(Q_{S,T}, K_{S,T}))}{T \ni Q_{S,T}, T \ni H(Q_{S,T}, K_{S,T})}, \dfrac{T \ni (Q_{seq}, Q_{ID})}{T \ni Q_{seq}}$

Similarly to (v), we infer that $T$ possesses the rest of the fields of $Q_{S,T}$. From (i) and (v), using the simple mechanism explained in D.3, $T$ verifies that $Q_{seq} \notin N^{p}_{S,T}$. Consequently, (vi) $T \mid\equiv \#(Q_{seq})$, $T$ believes that the message (including the fields that are omitted here) is fresh. Then, from (i), (iii), (iv), (vi), we get:

(vii) $$\frac{T \triangleleft H(Q_{S,T}, K_{S,T}), T \ni (Q_{S,T}, K_{S,T}), T \mid\equiv S \xleftrightarrow{K_{S,T}} T,\quad T \mid\equiv \#(Q_{S,T}, K_{S,T})}{T \mid\equiv S \mid\sim (Q_{S,T}, K_{S,T}), T \mid\equiv S \mid\sim H(Q_{S,T}, K_{S,T})}$$

This signifies the belief that both the packet payload and the *MAC* originate from $S$. Along with freshness (vi), we have the sought goal satisfied. We should note that the last inference does not imply that the sender revealed the shared key. In fact, the confirmation is independent of this issue. Moreover, we have assumed that none of the two principals compromises the shared secret by exposing it.[7]

Similarly, for message (2), we get:

(viii) $\dfrac{S \ni (R_{S,T}, H(R_{S,T}, K_{S,T}))}{S \ni R_{S,T}, S \ni H(R_{S,T}, K_{S,T})}, \dfrac{S \ni (Q_{seq}, Q_{ID}, route)}{S \ni Q_{seq}}$

(ix) $\dfrac{S \mid\equiv \#(Q_{seq})}{S \mid\equiv \#(Q_{seq}, route)}$

And finally

(x) $$\frac{S \triangleleft H(R_{S,T}, K_{S,T}), S \ni (R_{S,T}, K_{S,T}), S \mid\equiv S \xleftrightarrow{K_{S,T}} T,\quad S \mid\equiv \#(R_{S,T}, K_{S,T})}{S \mid\equiv T \mid\sim (R_{S,T}, K_{S,T}), S \mid\equiv T \mid\sim H(R_{S,T}, K_{S,T})}$$

Accordingly, $S$ believes that the entire route reply datagram originates from $T$ and is fresh and, trivially, that $T$ has constructed *route*, i.e., the source-route of the reply packet. The assumption of the non-colluding nodes implies that there is no alternative way for the route reply to arrive, but the one defined in the source-route. Moreover, the reply is the path along which the route request had propagated, which implies that the reply content had not been manipulated prior to its construction by $T$. Thus, its arrival at $S$ implies that the corresponding connectivity information is correct.

By updating the state at both ends, we can repeat the above reasoning to conclude that, if the source increments $Q_{seq}$ and does not repeat it within the lifetime of a *SA*, the sought goals are achieved, including the preservation of message integrity. In a very similar manner, this conclusion can be reached for the case of replies generated by intermediate nodes, under the assumption that the route suffix will be correct.[8]

[6] Due to the fixed-size headers fields and the well-defined structure of control traffic packets

[7] $K_{S,T}$ is used solely during the lifetime of the *S,T SA*.

## F. DISCUSSION

An interesting characteristic of the proposed here protocol is that it is essentially immune to *IP* spoofing. Any intermediate node may use any arbitrary *IP* address when queried but, as shown by the previous discussion, the protocol is capable of capturing the correct and current connectivity snapshot. However, in practice, neighbor discovery that maintains information on the binding of the *Medium Access Control* and *IP* addresses of nodes can strengthen the protocol. For example, the priority mechanism (D.2) would become more effective if packets were discarded when relayed by the same data link interface, i.e., the same Medium Access Control address, with more than one different *IP* addresses. Then, a malicious node would not be able to forge different *IP* address in different packets it relays, or, in other words, mask its misbehavior by appearing as a number of different nodes, and thus avoid being delegated to a lower priority.

Nevertheless, the issue of fair utilization of the network resources and possible ways to dismay nodes from broadcasting at the highest possible rate is beyond the scope of the security of routing protocols. For example, a malicious node could simply use *IP* broadcast instead of the route discovery querying mechanism. It is important though to defend nodes from attacks that exploit the protocol itself, and *SRP* provides protection against clogging *DoS* attacks. The replay protection at the end nodes, the use of the computationally inexpensive *HMAC* and the avoidance, in general, of any cryptographic validation by intermediate nodes are such features. These features are complemented by the scheme that regulates the propagation of queries. As a thought for future work, it would be interesting to investigate whether the use of soft state at intermediate nodes would further contribute to the protocol efficiency in a non-benign environment.

Moreover, it is important that the application of *SRP* does not severely affect the efficiency of the basis protocol under benign conditions. On one hand, in the same *MANET* subnet, nodes that implement *SRP* can co-exist with nodes that do not. In the absence of adversaries, the only overhead would be imposed on the nodes executing *SRP*. On the other hand, possible optimizations incorporated into the basics protocol can retain the effectiveness of the protocol in conjunction with *SRP*; an example is route shortening [8] that can be applied during the query propagation phase, based on knowledge of an active route. Finally, the fixed transmission overhead of *24* (or *27*) bytes per control packet becomes less significant as wireless network speeds increase to above the current state-of-the-art of *11Mbps*.

As shown above, the basic form of *SRP* that requires the propagation of queries to the destination is robust to malicious behavior. It is noteworthy that this statement remains true, in the absence of collusion, even if the destination node attempted to provide false replies. On the other hand, the provision of replies from intermediate nodes can achieve the same level of assurance only if a trusted node is assumed to provide a correct route segment. The reason is that even if some type of *HMAC* or signature from *T* were placed in the reply by *V*, it would still be possible for a stale $V \rightarrow T$ route segment to be provided, given that *S* cannot be assured of the $V \leftrightarrow T$ state. In practice, the usage of the *SA*-specific key (i.e., $K_{S,V}$) for such replies limits the effects of this potential residual vulnerability.

## G. CONCLUSIONS

In this paper, we proposed an efficient secure routing protocol for mobile ad hoc networks that guarantees the discovery of correct connectivity information over an unknown network, in the presence of malicious nodes. The protocol introduces a set of features, such as the requirement that the query verifiably arrives at the destination, the explicit binding of network and routing layer functionality, the consequent verifiable return of the query response over the reverse of the query propagation route, the acceptance of route error messages only when generated by nodes on the actual route, the query/reply identification by a dual identifier, the replay protection of the source and destination nodes and the regulation of the query propagation.

The resultant protocol is capable of operating without the existence of an on-line certification authority or the complete knowledge of keys of all network nodes. Its sole requirement is that any two nodes that wish to communicate securely can simply establish a priori a shared secret, to be used by their routing protocol modules. Moreover, the correctness of the protocol is retained irrespective of any permanent binding of nodes to *IP* addresses, a feature of increased importance for the open, dynamic, and cooperative *MANET* environments.

## APPENDIX

The basic notation used in E. is provided here, as in [26]. *X* and *Y* are formulas, *P* and *Q* are principals, *K* is a shared secret and *C* is a statement.

- *(X,Y): conjunction* of two formulas; it is treated as a set with properties of associativity and commutativity.

[8] The only way to avoid this assumption is to force the query to reach the destination. See section F. for a discussion on the trade-offs of this type of operation.

- **X*: *Not-originated-here* formula property. If *P* is told *X* (see below), it can distinguish it did not previously convey *X* in the current run.
- *H(X):* a *one-way function* of *X*.

Basic Statements

- $P \lhd X$ *: P is told* formula *X*.
- $P \ni X$ *: P possesses* or is capable of possessing formula *X*.
- $P \mid\sim X$ *: P once conveyed* formula *X*.
- $P \mid\equiv \#(X)$*: P believes* or is entitled to believe that formula *X* is *fresh*.
- $P \mid\equiv \varphi(X)$*: P believes* or is entitled to believe that formula *X* is *recognizable*, that is, *P* has certain expectations about the contents of *X* before actually receiving it.
- $P \mid\equiv P \stackrel{K}{\longleftrightarrow} Q$ *:* P *believes* or is entitled to believe that *K* is a *suitable secret* for *P* and *Q*.
- $C_1, C_2$*: conjunction* of two statements, treated as a set with properties of associativity and commutativity.
- $P \mid\equiv C$ *:* P *believes* or is entitled to believe that *statement C holds*.
- The *horizontal line* separating two statements or conjunctions of statements signifies that the upper statement *implies* the lower one. For example, $\frac{P \lhd (X,Y)}{P \lhd X}$ reads: *P being told* a formula implies *P being told* each of the formula's concatenated components.